\documentclass[twocolumn,prl,showpacs,superscriptaddress]{revtex4}

\usepackage{graphicx}
\usepackage{amsmath}
\usepackage{amssymb}
\usepackage[ansinew]{inputenc}
\usepackage{bm}\renewcommand{\bf}{\bm}
\def\vr{\bm{r}}
\newcommand{\pd}{\phantom{\dagger}}
\newcommand{\figwidth}{\columnwidth}

\begin{document}

\title{Competition between Vortex Unbinding \\
and  Tunneling in an Optical Lattice}

\author{M.~A. Cazalilla}

\affiliation{Donostia International Physics Center (DIPC),
Manuel de Lardizabal 4, 20018-Donostia, Spain.}

\author{A. Iucci}

\author{T. Giamarchi}

\affiliation{DPMC-MaNEP, University of Geneva, 24 Quai Ernest Ansermet
CH-1211 Geneva 4, Switzerland.}

\date{\today}

\begin{abstract}
We study a system of two-dimensional Bose gases
trapped in minima of a deep one-dimensional optical lattice potential.
Increasing the tunneling amplitude between adjacent gases
drives a deconfinement transition to a phase where
coherence is established between neighboring two-dimensional gases.
We compute the signature of this transition in the interference pattern
of the system as well as in its rotational response, which provides a direct 
measurement of the superfluidity in the system. 
\end{abstract}
\maketitle

A recent experiment by the ENS group \cite{hadzibabic_ENS_experiment_BKT}
has provided fairly direct evidence for the mechanism driving
the remarkable phenomenon known as
Berezinsky-Kosterlitz-Thouless (BKT) transition. The latter was
predicted~\cite{berezinsky_nolongrange_2d,kosterlitz_thouless,kosterlitz_renormalisation_xy}
in the 1970s and, in an two-dimensional (2D)  interacting Bose gas,
it is caused by  the proliferation of  (thermally excited) unbound pairs 
of vortices and anti-vortices  above a  temperature,
$T_{BKT}$, which  depends on the microscopic parameters of the
gas. As the transition takes place in a 2D system, long range
order is absent~\cite{mermin_wagner_theorem} both above and
below $T_{BKT}$  because of
the enhanced effect of thermal fluctuations.
Thus, the BKT transition, albeit continuous,  is not
characterized by the  emergence of an order parameter,
and therefore does not fit into the general
theory for phase transitions established by Landau in the 1930s.

In the ENS experiment~\cite{hadzibabic_ENS_experiment_BKT}
two independent, harmonically trapped,  2D Bose gases
were created after loading a (several tens of nK) cold
$^{87}$Rb gas  in the standing wave formed by two 
counter-propagating laser beams (\emph{i.e.} a 1D optical lattice). Since each 2D  gas is a finite-size
pancake, what was observed is more a sharp BKT
crossover than a phase transition. However,
by direct imaging of the expanded gas clouds after release from
the trap,  it was possible to show that, at  high enough temperature, 
the 2D Bose gases contain free vortices that leave a distinctive signature in the
interference pattern observed in time-of-flight experiments.
This provided sufficient direct evidence for the BKT
mechanism, which could not be obtained in other systems, such
like liquid helium films, where the BKT transition was observed
much earlier \cite{bishop_BKT_helium}.

Concerning the possibility of observing this transition in trapped
ultracold atomic gases,  there has been much debate in the
literature~\cite{tlho_low_d_bosons,petrov_BEC_in_quasi2D,alkhawaja_Stoof_lowD_gases,holzmann_KT_in_2D_gases,
simula_on_KT_trapped_gases}. One of the issues requiring
detailed investigation was that, being the
experimental systems harmonically trapped, they are inhomogeneous.
This, for instance, causes a non-interacting 2D Bose gas to become a true
Bose-Einstein condensate
(BEC)~\cite{petrov_BEC_in_quasi2D,holzmann_KT_in_2D_gases},  something that  is
not possible for  the homogeneous 2D
system~\cite{mermin_wagner_theorem,holzmann_KT_in_2D_gases}. However,
recent theoretical calculations~\cite{simula_on_KT_trapped_gases}
have clarified this issue showing that, in spite of  the
inhomogeneity, thermal activation of vortex-anti-vortex  pairs, and
thus a BKT-like crossover, can take place also in the trapped
systems. This  was later confirmed by the
experiments~\cite{hadzibabic_ENS_experiment_BKT}.

As we show in this paper, 1D optical lattices, like
the one used in \cite{hadzibabic_ENS_experiment_BKT}, offer the possibility of
exploring a much richer kind of phenomena, which lies beyond the
standard BKT physics: due to the large degree of
tunability of these  atomic systems, it is
experimentally possible to control the tunnelling amplitude between the
pancakes that are formed in the deep wells of the 1D lattice. This
adds a new dimension  to the phase diagram of the system:
the Josephson coupling between the 2D gases, which indeed can be
more easily controlled than temperature.  When the amplitude of
this coupling, $t_{\perp}$, is made large enough so that tunneling of
an appreciable number of  bosons takes place over the typical
duration, $t_{\rm exp}$, of an experiment  (\emph{i.e.}
$\hbar/t_{\rm exp} \ll  t_{\perp}$), but small enough so that the
pancakes remain 2D (\emph{i.e.} $t_{\perp} \ll \mu$, $\mu$ being
the chemical potential of the 2D gas), we find that the Josephson
coupling can drive a transition where phase coherence builds up
between neighboring pancakes. The competition between the interplane phase coherence and the 
thermally excited vortices in the pancakes leads to a transition analogous 
to the \emph{deconfinement} transition studied by some of
us~\cite{donohue_commensurate_bosonic_ladder,ho_deconfinement_coldatoms,cazalilla_deconfinement_longpaper}
in various types of coupled quantum 1D interacting  Bose systems, such
like those that are created in anisotropic optical
lattices~\cite{stoferle_tonks_optical}. Universality between classical
statistical mechanics in 2D and quantum field  theories in 1D allows to use 
the same techniques \cite{donohue_commensurate_bosonic_ladder,cazalilla_deconfinement_longpaper,benfatto_BKT_in_cuprates,mathey_twoplanes_cold} to tackle the coupled classical two dimensional systems as well. In addition to their intrinsic interest, 
the physics of coupled 2D Bose systems is also relevant
to the theoretical understanding of recent experiments in the
high-$T_c$ cuprate superconductors (see~\cite{benfatto_BKT_in_cuprates} and
references therein), for which control over the system parameters
is much more limited.

\begin{figure}
\begin{center}
\includegraphics[width=\figwidth]{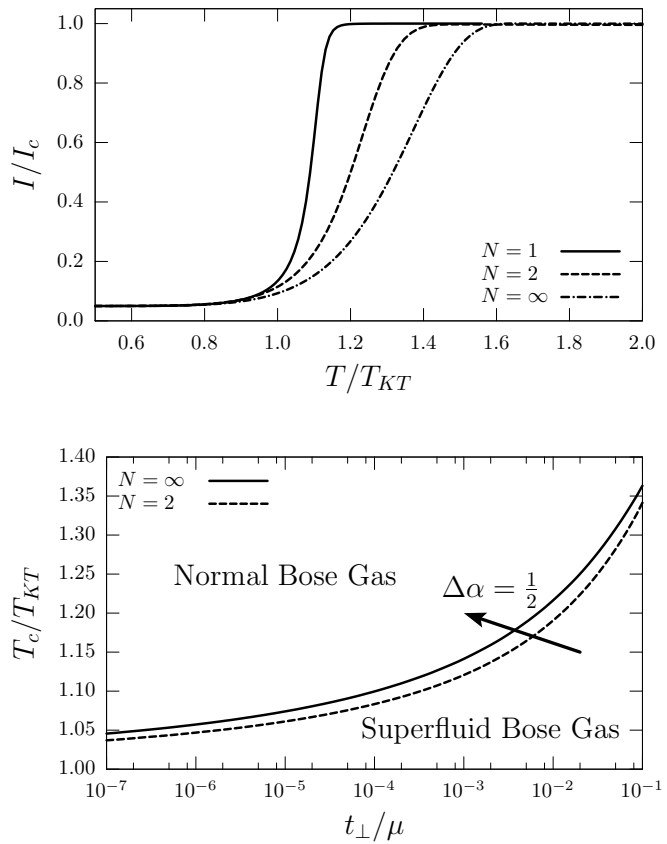}
\caption{Upper panel: Ratio of the moment of inertia to the
classical moment of interia: $I/I_{\rm cl} = (1 -
\rho_s/\rho)$ for $N = 2$ and $N \to \infty$; $t_{\perp} = 0.05\, \mu$.
To mimic finite size effects we have assumed that the 1D lattice is
made of circular pancake-shaped 2D Bose gases of radius $R  = 100 \:
\xi$, where $\xi$ is the healing length, and therefore we have stopped the
renormalization-group flow at the scale of $R$ (furthermore, we have assumed that $\rho_0/\rho = 0.95$). Lower panel: Phase diagram (in the thermodynamic limit). We have indicated with an arrow the jump in the exponent ($\alpha$)
that describes the scaling of the interference contrast with the imaged length $L_x$ (see Eqs.(\ref{eq:contrast},\ref{eq:pcontrast})).}\label{fig:fig1}
\end{center}
\end{figure}

As shown in what  follows, when only two pancakes are coupled, the resulting
system behaves, below a certain temperature
$T_c$ (see lower panel of Fig.~\ref{fig:fig1}),  as single 2D 
superfluid. Furthermore, if an infinite
number of pancakes are coupled, the deconfinement
transition yields a true BEC. Here we focus on the experimental
consequences of this transition: although in the BEC or superfluid
phases the system displays true interference fringes,
a measurement of the interference contrast like the one conducted by
the ENS group will yield a jump, $\Delta \alpha = \frac{1}{2}$, in
the exponent ($\alpha$) that characterizes the scaling behavior of
the interference contrast with the imaged area ($\Delta \alpha =
\frac{1}{4}$ for the BKT transition~\cite{polkovnikov_interference_between_condensates,hadzibabic_ENS_experiment_BKT}). 
However, if the interference pattern serves as probe of the coherence between the pancakes, 
it is not a probe of  the superfluid properties of phase-coherent regime. We show that a 
\emph{direct} probe of the superfluidity is provided by measuring the
moment of inertia of  the system at low rotation frequencies.
In the presence of weak Josephson coupling between the
pancakes, the rotational response exhibits a behavior that differs from
that of a single (\emph{i.e.} uncoupled) pancake (see upper panel of
Fig.~\ref{fig:fig1}).

In the regime where  phase fluctuations are classical,
\emph{i.e.} for $T \gtrsim \mu$, the Bose field in the $m$-th ($m
= 1, \ldots, N$) pancake of a 1D lattice  can be  written (away
from the vortex cores) as $\Psi_m({\bf r})  = \rho^{1/2}_0 \:
e^{i \Theta_m({\bf r})}$ (${\bf r} = (x,y)$). Thus the 1D lattice is
described by the following Hamiltonian for a stack of $N$ coupled
XY models:
\begin{equation}
H = H_{\rm XY}
 -  \frac{g_{J}}{\pi a^2_0} \sum_{m=1}^{N-1}
 \int d{\bf r}\, \cos \left[  \Theta_m({\bf r}) -
\Theta_{m+1}({\bf r})\right] , \label{eq:coupledXY}
\end{equation}
where $ H_{\rm XY} =  \frac{K}{2\pi} \sum_{m=1}^{N}   \, \int
d{\bf r}\left( \nabla \Theta_m({\bf r}) \right)^2$ is the
Hamiltonian of the uncoupled pancakes of area $A$.  We have
assumed the pancakes to be  identical and, for most of what follows,
we shall take them to be infinite (\emph{i.e.} $A \to
\infty$). The \emph{bare} parameters are $K(0) = \pi\hbar^2
\rho_0/MT$, where $M$ is the boson mass, $T$ the temperature, and
$\rho_0 = \rho_0(T) = |\Psi_m({\bf r})|^2$, the quasi-condensate
density~\cite{alkhawaja_Stoof_lowD_gases,petrov_BEC_in_quasi2D}; 
the (dimensionless) strength of the Josephson coupling
$g_{J}(0) = \pi a^2_0\rho_0 t_{\perp}/T$, where $a_0$ is of the
order of the healing length $\xi = \sqrt{\hbar^2/M\mu}$.  It can
be shown  that, close to the BKT transition, the partition function
of (\ref{eq:coupledXY}) can be obtained from the  partition
function of a  1D \emph{quantum} model with the following
Hamiltonian~\cite{giamarchi_book_1d}:
\begin{eqnarray}
\hat{H} &=& \hat{H}_{\rm sG} - \frac{g_J}{\pi a^2_0} \sum_{m=1}^{N-1} \int
dx
\cos \left[ \theta_{m}- \theta_{m+1} \right], \label{eq:coupledDsG}\\
\hat{H}_{\rm sG} &=& \sum_{m=1}^{N} \int  \frac{dx}{2\pi} \,   \Big[
 K \left(\partial_x \theta_m \right)^2  \nonumber  + K^{-1}
 \left(\partial_x \phi_m \right)^2 \nonumber \\
&& \quad - \frac{2 g_v}{a^2_0} \cos 2 \phi_m \Big].
\label{eq:sG}
\end{eqnarray}
The operators $\theta_m$ and $\partial_x
\phi_m$ are canonically conjugated fields. The coupling $g_v =
2\pi e^{-E_C/T}$, where $E_C \simeq 1.6\pi \: \hbar^2
\rho_0(T)/M$ is the vortex core energy~\cite{minnhagen_vortex_core_energy,benfatto_BKT_in_cuprates}.

To ascertain the nature of the phases of the 1D optical
lattice, we apply the renormalization group (RG) to (\ref{eq:coupledDsG}), 
\emph{i.e.} we iteratively 
coarse-grain short-distance degrees of freedom to obtain a simpler description of the system. To analyze the dependence on the number of coupled pancakes, we consider two limiting cases: the double pancake system ($N  = 2$) and the infinite lattice
($N= \infty$). We show that the main features of the transition in both
cases are similar.  In the calculations, we have
exploited that (\ref{eq:coupledDsG}) also describes a system of
Josephson-coupled quantum 1D  Bose gases in a periodic potential at
commensurate filling~\cite{donohue_commensurate_bosonic_ladder,ho_deconfinement_coldatoms,
cazalilla_deconfinement_longpaper}. More details on the RG procedure can be found in 
\cite{cazalilla_deconfinement_longpaper,giamarchi_book_1d}.

Let us start with the two pancake system. We
introduce symmetric and anti-symmetric field
combinations \cite{donohue_commensurate_bosonic_ladder}
$\phi_{\pm} = \left(\phi_1 \pm \phi_2 \right)/\sqrt{2}$ and
$\theta_{\pm} = \left( \theta_{1}  \pm \theta_2 \right)/\sqrt{2}$.
The action for these fields depends on the parameters $K_{\pm},
g_{v}$ and $g_J$, which flow under the RG transformations
as described by the following set of
equations \cite{donohue_commensurate_bosonic_ladder,mathey_twoplanes_cold}:
\begin{equation}\label{eq:rg21} 
\begin{split}
\frac{d K_{+}}{d\ell} =  -g^2_v K^2_{+}\quad\quad \frac{dK_{-}}{d\ell} =
- g^2_v K^2_{-} + \frac{g^2_J}{2}, \\
\frac{d g_v}{d\ell} = \left[ 2 - \frac{K_{+} + K_{-} }{2} \right] g_v,
\,\frac{d g_J}{d\ell} = \left( 2 - \frac{1}{2K_{-}}\right) g_J. 
\end{split}
\end{equation}
The above equations describe the crossover between two strong
coupling RG fixed points: when the bare couplings determined by
$T, \rho_0,t_{\perp}$, and $M$, are such that  $g_v$
becomes of order one first (and the perturbative method employed
to obtain (\ref{eq:rg21}) breaks down), vortex
unbinding takes place before phase coherence between the two
pancakes is established. In this phase the two pancakes are
decoupled and behave as normal Bose gases. On the
other hand, when the bare couplings are such that $g_J$ becomes of
order one first, the Josephson coupling term ($\propto  g_J$)
causes the relative phase $\theta_{-}$ of the two pancakes to lock
to the same value (modulo $2\pi$). In this (superfluid) phase, only the
symmetric field $\theta_{+}$ can fluctuate at large
distances. This causes the same decay of
 correlations in both pancakes:  $\langle
\psi_{m}({\bf r}) \psi_{m}({\bf 0}) \rangle \sim |{\bf
r}|^{-1/2K_{+}}$  ($m = 1,2$).  We note that $K_{+}$ is
related to the superfluid density,  $\rho_s(T)$. The relationship is obtained from
the transformation of~(\ref{eq:coupledDsG})
under a phase twist where $\partial_x\theta_m \to \partial_x\theta_m + \alpha$
(\emph{i.e.} $\partial_x \theta_{+} \to \partial_x \theta_{+} + \alpha \sqrt{2}$ but
$\theta_- \to \theta_-$). Thus,
$K_+/\pi = \hbar^2 \rho_s(T)/M T$.

The case of large number of pancakes $N\to \infty$ can be treated by a similar application of the
RG to (\ref{eq:coupledDsG}). This leads the following set of flow equations~\cite{ho_deconfinement_coldatoms,cazalilla_deconfinement_longpaper,benfatto_BKT_in_cuprates}:
\begin{equation}\label{eq:rgi1}
\begin{split}
\frac{dK}{d\ell} = 2 g^2_J - K^2 g_v, \quad
\frac{dK_s}{d\ell} =  -g^2_v K^2_s,  \\
\frac{dg_v}{d\ell} = (2 - K) g_v, \, \frac{dg_J}{d\ell} = \left(2
- \frac{1}{4K} - \frac{K_s}{4K^2}  \right) g_J.
\end{split}
\end{equation}
The physics described by these equations is similar: as we coarse grain the short distance degrees of
freedom of the system, $g_J$ and $g_v$ get renormalized.  For
given  values of $T$, $t_{\perp}$, $\rho_0$, and $M$, whichever $g_v$
or $g_J$ reaches a value of order one first, determines the phase
of the system. When the coupling $g_v$ dominates, vortex unbinding
destroys  phase coherence within the pancakes and therefore
suppresses Josephson tunnelling at long distances: the pancakes
decouple and an array of normal gases is obtained. On the other
hand, if $g_J$ dominates over $g_v$,  phase coherence builds up
and a BEC forms. The resulting system is an anisotropic  3D BEC,
whose properties can be computed by using Landau-Ginzburg
theory~\cite{cazalilla_deconfinement_longpaper}. By the same `phase-twist'
argument given above, it is found that $K_s/\pi = \hbar^2
\rho_s(T)/MT$.

Using the above RG equations we can compute the observable consequences of the 
deconfinement transition. Let us first discuss the interference contrast that was
measured by the ENS group~\cite{hadzibabic_ENS_experiment_BKT,polkovnikov_interference_between_condensates}.
It was argued in Ref.~\cite{polkovnikov_interference_between_condensates} that
interference between two independent \emph{identical} Bose systems
provides a way to probe correlations in each system. This idea was
employed by the ENS group to observe the BKT transition by
creating a double-pancake system in a 1D optical lattice and making
the atom hopping between them negligible over the time scale of
the duration of the experiment.  However, as discussed above, when
hopping between pancakes is increased, it leads to a phase where
they are \emph{not} independent anymore. Thus, the expansion
images taken after release of the atom clouds from the traps
will exhibit \emph{true} interference fringes.  For instance,
for two pancakes in the superfluid phase,
the imaged (integrated) density distribution of the expanded coulds
will exhibit oscillations that do not wash out  after averaging over many
experimental runs.  On the other hand,
in the normal gas phase (as well as in the superfluid phase at $t_{\perp} =0$) where the pancakes are independent,  
these oscillations average to  zero.
However, if while seeking for signatures of the  transition between these two phases, one insists in measuring the interference contrast, 
we show below that the latter also provides information about the phase correlations between
neighboring 2D gases.

For $N = 2$, following~\cite{polkovnikov_interference_between_condensates}, we
consider the integrated interference contrast,
\begin{equation}\label{eq:contrast}
\langle\vert C_Q\vert^2\rangle= (L_x L_y)^{-2}
\int_{L_x \times L_y} d\vr_1
d\vr_2 F_{12}({\bf r}_1, {\bf r}_2),
\end{equation}
where $F_{12}({\bf r}_1,{\bf r}_2) = \langle\Psi^{\dagger}_2(\vr_1)\Psi^{\pd}_1(\vr_1) \Psi^{\dagger}_1(\vr_2)\Psi^{\pd}_2(\vr_2)\rangle$;
$\langle\vert C_Q\vert^2\rangle$ is the square of the amplitude of the  oscillations in the interference pattern  with wave number $Q=md/t$, normalized to (the square of)
the imaging length $L_x$ ($L_y$ is kept fixed)~\cite{hadzibabic_ENS_experiment_BKT, polkovnikov_interference_between_condensates}. In terms of symmetric and antisymmetric fields,  $F_{12}({\bf r}_1,{\bf r}_2) \propto \langle
e^{i\sqrt{2}\theta_-(\vr_1)}e^{-i\sqrt{2}\theta_-(\vr_2)}\rangle$.
In the normal Bose gas phase, the relative phase $\theta_{-}$
undergoes wild fluctuations because vortex-anti-vortex pairs unbind.
Thus correlations of $\theta_-$  asymptotically decay to zero exponentially.
This implies that $\langle\vert C_Q\vert^2\rangle\sim L_x^{-2\alpha}$ with $\alpha=\frac{1}{2}$~\cite{polkovnikov_interference_between_condensates}. On the other hand, when the Josephson coupling dominates, long-distance behavior of the system, the energy of the system is minimized by locking $\theta_-$  into one of the minima of the cosine in the second term of~(\ref{eq:coupledDsG}).
Hence, in the $\vert\vr_1  - \vr_2\vert\to\infty$ limit, $\langle
e^{i\sqrt{2}\theta_-(\vr_1)}e^{-i\sqrt{2}\theta_-(\vr_2)}\rangle \to \langle
e^{i\sqrt{2}\theta_-}\rangle\langle
e^{-i\sqrt{2}\theta_-}\rangle = $ const. Therefore, in the superfluid
phase $\alpha=0$. Thus, in going
from the superfluid phase to a phase of decoupled 2D normal Bose
gases by changing $T$ (or, easier, $t_\perp/\mu$), we
predict a stronger change in  the scaling behavior of the interference
contrast: $\langle\vert
C_Q\vert^2\rangle\sim L_x^{-2\alpha}$, where $\alpha$ changes from
$0$ in the superfluid to $\frac{1}{2}$ for the normal gas phase.
Notice that this is a much larger jump than the one
predicted~\cite{polkovnikov_interference_between_condensates} (and
observed \cite{hadzibabic_ENS_experiment_BKT}) for the standard BKT transition
in a 2D Bose gas, where $\alpha$ jumps from $\alpha = \frac{1}{4}$
at $T < T_{KT}$ to $\alpha = \frac{1}{2}$ at $T>T_{KT}$.

For $N = \infty$, the interference pattern receives contributions 
from an infinite number pancakes. Thus, the correlation function of the
integrated density after expansion reads:
\begin{equation}
\langle\rho(z_1)\rho(z_2)\rangle=\sum_p
e^{ipQ(z_2-z_1)}N^2 (L_x L_y)^2\: \langle\vert C_Q^p\vert^2\rangle,
\end{equation}
where $z$  is the coordinate of along the axis normal to the pancakes and
 the $p$-th contrast is
\begin{multline}
\langle\vert C_Q^p\vert^2\rangle=\frac{(L_x L_y)^{-2}}{N^2}\sum_{mn}
\int_{L_x\times L_y}
d\vr_1 d\vr_2\,e^{ipQ(m-n)d}\\
\langle\Psi^{\dagger}_{m+p}(\vr_1)\Psi^{\pd}_m(\vr_1)
\Psi^{\dagger}_n(\vr_2)\Psi^{\pd}_{n+p}(\vr_2)\rangle.\label{eq:pcontrast}
\end{multline}
Nevertheless, in the (superfluid)  BEC  phase, the above
correlation functions also decay to a constant asymptotically
for $|{\vr_1}-{\vr_2}| \to \infty$. Therefore, $ \langle\vert
C_Q^p\vert^2\rangle\sim L_x^{2\alpha}$ with $\alpha =0$. In the
normal gas phase, the pancakes decouple, and one finds  $
\langle\vert C_Q^p\vert^2\rangle\sim L_x^{2\alpha}$ with $\alpha
=\frac{1}{2}$. Thus the same jump in $\alpha$ as for $N=2$ should
be observed.

Whereas the interference contrast
provides evidence for the existence or lack of phase
coherence between pancakes, it is not a direct measurement of the
superfluid density $\rho_s$. Such a measurement
can obtained from the response of the system
to a slow rotation about the axis of the 1D lattice ($z$-axis). In
the (rotating) reference frame where the gas is stationary, the
equilibrium state is found by minimizing the free energy $F(
\Omega) = - T \: \ln Z[ \Omega]$,  $Z[\Omega] = {\rm Tr} \,
e^{-(H - \Omega L_z)/T}$ with $H$ given by (\ref{eq:coupledXY})
and $L_z = \sum_{i}
\sum_{\alpha\beta}{\epsilon_{\alpha\beta}} r_{\alpha i}p_{\beta i}
$ being the $z$ component of the total angular momentum operator
(${\bf r}_i, {\bf p}_i $ are the position and canonical momentum
operators for the $i$-th particle, respectively), and
$\epsilon_{xy} = - \epsilon_{yx} = +1$.  By treating the term
$\Omega L_z$ as a perturbation (assuming $\Omega$ to be small),
we can write  $\Omega L_z = \int d{\bf r}\, {\bf
A}^{\Omega}(\bf r) \cdot {\bf j}({\bf r})$, ${\bf j}({\bf r}) =
\frac{\hbar}{2i} \sum_{m=1}^N \left[\Psi^{\dag}_m({\bf r})  \nabla
\Psi_m({\bf r}) - \nabla \Psi^{\dag}_m({\bf r}) \Psi_m({\bf r})
\right]$ being the  momentum density and ${\bf A}^{\Omega}({\bf
r}) = \Omega (-y, x)$ an effective \emph{transverse}  (\emph{i.e.}
$\nabla \cdot {\bf A}^{\Omega}({\bf r}) = 0$) vector potential.
Within linear response theory, the angular momentum of the lattice
$\langle L_z  \rangle_{\Omega \neq 0} = \sum_{\alpha \beta}\int
d{\bf r} \,  \epsilon_{\alpha \beta} \, r_{\alpha} \: \langle
j_{\beta}({\bf r}) \rangle_{\Omega \neq  0 }  = \sum_{\alpha \beta
\gamma}  \int d{\bf r} d{\bf r'}\, \epsilon_{\alpha \beta}\,
r_{\beta}  \,  C_{\alpha \beta}({\bf r},{\bf r}')
A^{\Omega}_{\gamma}({\bf r}') + O(\Omega^2)$. Thus the response to
rotation is given by the momentum density correlation function at
zero rotation frequency, $C_{\alpha\beta}({\bf r},{\bf r}') =
\frac{1}{T} \langle j_{\alpha}({\bf r}) j_{\beta}({\bf r}'
)\rangle_{\Omega = 0}$. The latter can be decomposed into a
longitudinal and a transverse part,
$C_{\alpha \beta}({\bf q}) =  F_{||}(q) \hat{q}_{\alpha}
\hat{q}_{\beta} + F_{\perp}(q) \left( \delta_{\alpha \beta} -
\hat{q}_{\alpha} \hat{q}_{\beta}\right)$, in Fourier space, where
$\hat{q}_{\alpha} = q_{\alpha}/q$, and $F_{||}(q)$ and
$F_{\perp}(q)$ are functions of $q = |{\bf q}|$ only. Since ${\bf
q}\cdot {\bf A}^{\Omega}({\bf r}) = 0$,  only the transverse part
of the correlation function, which is associated with the
irrotational part of the gas flow caused by the
vortices~\cite{chaikin_lubensky}, will contribute to the response
to rotation. For small $\Omega$, the vector potential is a
sufficiently slow perturbation so that only the $q = 0$
component of the correlation function matters, and it can be
shown~\cite{chaikin_lubensky} that $F_{\perp}(0) = \rho_n = \rho -
\rho_s$, where $\rho$ is the gas density. Hence, $\langle L_z
\rangle_{\Omega \neq 0} = I\: \Omega \simeq (1  - \rho_s/\rho)
I_{\rm cl} \Omega$, where $I_{\rm cl}$ is the classical moment of
inertia of the pancake.  Since  $\rho_s$ is
related to $K_{+}$ ($N  = 2$) or $K_s$ ($N \to \infty$),
 the behavior of the ratio $I/I_{\rm cl} = (1 -
\rho_s/\rho)$ can be obtained from the RG equations~(\ref{eq:rg21},\ref{eq:rgi1}).
For $N = 2$,  in the normal gas phase, $K_{+} \to
0$ ($K_s \to 0$, for $N = \infty$) and  $I/I_{\rm cl} \to 1$.
On the other hand, in the superfluid phase, $K_{+}$ ($K_{s}$) flows to a
finite value and therefore $I/I_{\rm cl} < 1$ because the superfluid
fraction does not respond to rotation. The transition between these two
phases is shown in the upper panel of Fig.~\ref{fig:fig1}.
The (abrupt, for $N =1$)  change in $I/I_c$ between the superfluid
(or BEC) and the normal gas phases  is rendered smooth(er) by
finite-size effects, which are mimicked by stopping the RG flow at a length
scale equal to the size of the pancake ($R = 100 \: \xi$ in Fig.~\ref{fig:fig1}).
The change may be further smoothen by inhomogeneity effects~\cite{holzmann_KT_in_2D_gases}. Experimental determination of $I/I_{\rm cl}$ by measuring
$\langle L_z \rangle_{\Omega \neq 0}$ is not an easy task. Nevertheless,
obtaining information on $\langle L_z \rangle_{\Omega}$, and hence
on $I/I_{\rm cl}$, is possible by measuring the shape oscillations
of the rotating pancakes, the so-called  `scisors mode'
oscillation caused by a sudden rotation
of a quadrupole deformation of the trap~\cite{zapelli_moment_of_inertial_BEC}.

We thank L. Benfatto for many enlightening discussions. This work
was supported by \emph{Gipuzkoako Foru Aldundia} and MEC (Spain)
under grant  FIS2004-06490-C03-00 (MAC) and the Swiss National
Science Foundation under MaNEP and Division II (AI and TG).

\bibliography{1DOL,totphys}

\end{document}